\documentclass[preprint,showpacs,amsfonts,amsmath,amssymb]{revtex4}
\usepackage{graphicx}% Include figure files
\usepackage{dcolumn}% Align table columns on decimal point
\usepackage{bm}% bold math

\begin{document}
%\preprint{example}
\title{Distribution of Entropy of Bardeen Regular Black Hole with
Corrected State Density }
%\altaffiliation{}
\author{Hai Huang}
\author{Juhua Chen} \email{jhchen@hunnu.edu.cn}
\author{Yongjiu Wang}
\affiliation{College of Physics and Information Science, Hunan
Normal University, Changsha, Hunan 410081, P. R. China}
\begin{abstract}
We consider corrections to all orders in the Planck length on the quantum state density, and calculate the statistical entropy of the scalar field on the background of the Bardeen regular black hole numerically. We obtain the distribution of entropy which is inside the horizon of black hole and the contribution of the vicinity of horizon takes a great part of the whole entropy.


\end{abstract}
\maketitle
In 1970's, Bekenstein and Hawking found that the black hole(BH) entropy was proportional to the area of the event horizon by comparing BH physics with thermodynamics\cite{1}.'t Hooft statistically obtained the Bekenstein-Hawking entropy of a scalar field outside the horizon of the Schwarzschild BH by brick-wall model\cite{2}. The Bekenstein-Hawking entropy was identified with the statistical-mechanical entropy arising from a thermal bath of quantum fields propagating outside the event horizon,but there are some drawbacks like ultraviolet divergence near the horizon. An improved brick-wall method has been introduced by taking the thin-layer outside the event horizon of a BH as the integral region, but the artificial cutoffs is still unsolved\cite{3}. It's commonly believes that there exists a minimal length which originates due to the quantum fluctuation of the gravitational field motivated by the generalized uncertainty principle (GUP) \cite{4,5}, the entropy integral of a radial component in the range near the horizon should be treated for a convergent entropy\cite{6}. The Feynman propagator displays an exponential ultraviolet cutoff of the form exp($-\lambda p^2$), where the parameter $\lambda$ actually plays the role of a minimal length\cite{7}. Quantum gravity phenomenology has been tackled with effective models based on GUPs and/or modified dispersion relations containing a minimal length as a natural ultraviolet cutoff \cite{7.1,7.2,7.3}. At the quantum mechanical level, the essence of the ultraviolet finiteness of the Feynman propagator can be captured by a nonlinear relation $p=f(k)$, where $p$ and $k$ are the momentum and the wave vector of the particle, respectively. The commutator between the operators $x_i$ and $p_j$ is generalized to $[x_i,p_j]=i\frac{\partial p_i}{\partial k_j}$; moreover, the usual momentum measure in $3+1$-dimensional spacetime $d^3p$ is modified to
\begin{equation}
d^3p\ \textrm{det}\left|\frac{\partial k_i}{\partial p_j}\right|=d^3p \prod_i\left|\frac{\partial k_i}{\partial p_j}\right|,
\end{equation}
where $\frac{\partial k_i}{\partial p_j}=\delta_{ij}e^{\lambda p_i^2}$\cite{8,9,10}, the number of quantum states in a volume element in phase cell space based on the GUP is
\begin{equation}
dg_+(\omega)=\frac{d^3xd^3p}{(2\pi\hbar)^3}e^{-\lambda p^2},\label{g+}
\end{equation}
where $p^2=p_ip^i$ is the square of momentum. Many authors calculated the entropy of BHs to the leading order in the Planck length by using this newly modified equation of states of density\cite{11,12,13,14}. In their works, the entropy were concentrated on a small vicinity of the BH horizon, we do not know what's the contribution of the interior of BH to the entropy.

In this letter, we devote to consider the contribution of the whole interior of BH by numerical method. We study the entropy of a scalar field on a Bardeen regular black hole(BRBH) backgrounds. Contrary to our general expectation, this study of a regular case is not so trivial in contrast to the previous result of the  non-regular case. By using the equation of states of density  motivated by GUP in quantum gravity, we calculate the quantum entropy of a massive scalar field numerically. We obtain the desired Bekenstein-Hawking entropy through the whole event horizon and little mass approximation satisfying the asymptotic property of the wave vector $k$ in the modified dispersion relation. We take the units as $G=\hbar=c=k_B\equiv1$.

In the context of gravity coupled to some form of matter, regular solutions are obtained from a prototypical action of the form
\begin{equation}
S_{act}=\frac{1}{16\pi}\int d^4x\sqrt{-g}(R-\pounds),
\end{equation}
where $g$ is the determinant of the metric $g_{\mu\nu}$, $R$ is the scalar curvature, and $\pounds$ represents the Lagrangian of the matter fields. In the context of a specific nonlinear electrodynamics with Lagrangian given by $\pounds=(6m/(2|\alpha| \alpha^2))(\sqrt{2\alpha^2 F}/(1+\sqrt{2\alpha^2 F}))^{5/2}$, the BRBH line element takes the form
\begin{equation}
ds^2=-f(r)dt^2+\frac{1}{f(r)}dr^2+r^2(d\theta^2+sin^2\theta d\varphi^2),\label{metric}
\end{equation}
note that the function $f(r)$ takes a particularly simple form
\begin{equation}
f(r)=1-\frac{2mr^2}{(r^2+\alpha^2)^{3/2}},
\end{equation}
where $(t,r,\theta,\phi)$ are the usual space-time spherical coordinates , $m$ and $\alpha$ stand for the mass and the monopole charge of a self-gravitating magnetic field of a non-linear electrodynamics source, respectively(seeing Fig.1).
Depending on the value of $\alpha$, the above regular solution may present two distinct, one,or no horizons. It reduces to the Schwarzschild solution for $\alpha=0$.
\begin{figure}
  \centering
  % Requires \usepackage{graphicx}
  \includegraphics{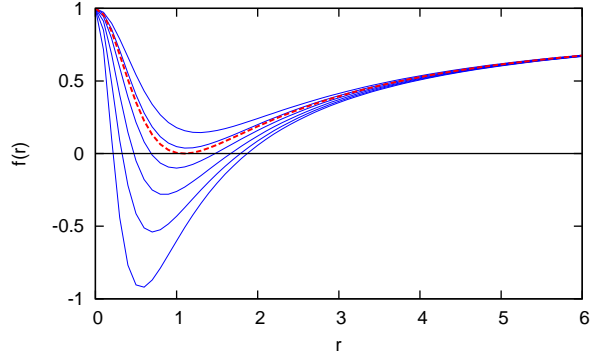}\\
  \caption{The figure illustrates the behavior of the metric function $f(r)$ given in (\ref{metric}) for different values of the parameter $\alpha$.}
\end{figure}
In this BRBH background, let us consider massless scalar field which satisfies the Klein-Gordon equation
\begin{equation}
\frac{1}{\sqrt{-g}}\partial_\mu(\sqrt{-g}g^{\mu\nu}\partial_\nu\Phi)=0.
\end{equation}
Substituting the line element(\ref{metric}) and using the leading order Wenzel-Kramers-Brillouin(WKB) approximation $\Phi=e^{-i\omega t}e^{iS(r,\theta,\phi)}$, Eq.(\ref{metric}) becomes
\begin{equation}
\frac{\omega^2}{f(r)}-f(r)\left(\frac{\partial S}{\partial r}\right)^2-\frac{1}{r^2}\left(\frac{\partial S}{\partial \theta}\right)^2
-\frac{1}{r^2sin^2\theta}\left(\frac{\partial S}{\partial \varphi}\right)^2=0.
\end{equation}
We have
\begin{equation}
p_r^2=\frac{1}{f(r)}\left(\frac{\omega^2}{f(r)}-\frac{1}{r^2}p_\theta^2-\frac{1}{r^2sin^2\theta}p_\varphi^2\right),
\end{equation}
where $p_r=\frac{\partial S}{\partial r}$, $p_\theta=\frac{\partial S}{\partial \theta}$ and $p_{\varphi}=\frac{\partial S}{\partial \varphi}$, we also obtain the square module momentum
\begin{equation}
p^2=p_ip^i=g^{r r}p_r^2+g^{\theta \theta}p_\theta^2+g^{\varphi \varphi}p_\varphi^2=\frac{\omega^2}{f(r)}.
\end{equation}
From Eq.(\ref{g+}), the number of quantum states with the energy is less than $\omega$ is given by

\begin{eqnarray}
% \nonumber to remove numbering (before each equation)
 g_+(\omega)&=&\frac{1}{(2\pi)^3}\int e^{-\lambda p^2}drd\theta d\varphi dp_rdp_\theta dp_\varphi  \nonumber \\
 &=&\frac{2}{(2\pi)^3}\int  e^{-\lambda p^2}drd\theta d\varphi \int \frac{1}{\sqrt{f(r)}}\left(\frac{\omega^2}{f(r)}-\frac{1}{r^2}p_\theta^2-\frac{1}{r^2sin^2 \theta} p_\varphi^2\right)^{\frac{1}{2}}dp_\theta dp_\varphi    \nonumber \\
  &=&\frac{2}{3\pi}\int\frac{r^2\omega^3}{f^2(r)}e^{-\lambda p^2}dr. \label{g}
\end{eqnarray}
By using Eq.(\ref{g}), the free energy can be rewritten as
\begin{eqnarray}
F_+&\approx&\frac{1}{\beta}\int dg_+(\omega)\ln(1-e^{-\beta \omega})  \nonumber \\
 &=& -\int\frac{g_+(\omega) d\omega}{e^{\beta \omega}-1}  \nonumber \\
 &=& -\frac{2}{3\pi}\int_0^{r_H} e^{-\frac{\lambda \omega^2}{f(r)}}\frac{r^2}{f^2(r)}dr\int_0^\infty \frac{\omega^3 d\omega}{e^{\beta \omega}-1}
\end{eqnarray}
Here, we have taken the continuum limit of quantum numbers and integrated it. Furthermore, the contribution from the vacuum surrounding the system is ignored, we are not only concerned about the contribution from the just vicinity near the horizon, but also the whole region within the horizon $(0\rightarrow r_H)$.

We can obtain the entropy as follows:
\begin{eqnarray}
% \nonumber to remove numbering (before each equation)
  S_+ &=& \beta^2 \frac{\partial F_+}{\partial \beta} \nonumber \\
   &=& \frac{2\beta^2}{3\pi}\int_0^{r_H}e^{-\frac{\lambda \omega^2}{f^2(r)}}\frac{r^2}{f^2(r)}dr\int_0^\infty \frac{e^{\beta \omega}\omega^4d\omega}{(e^{\beta \omega}-1)^2}
\end{eqnarray}
\begin{figure}
  \centering
  % Requires \usepackage{graphicx}
  \includegraphics{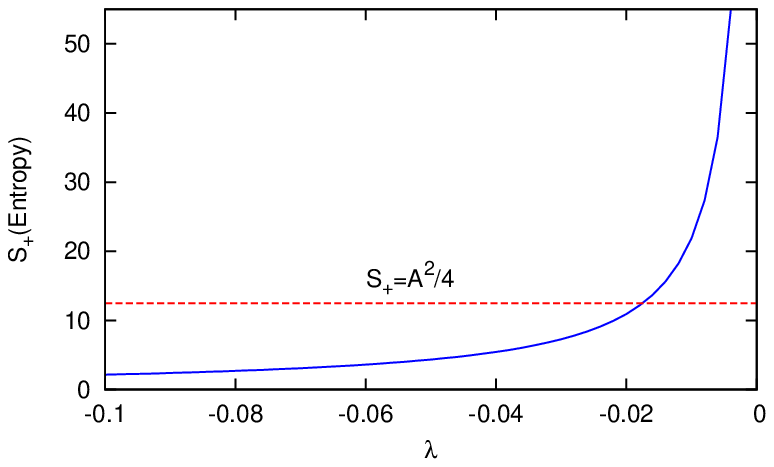}\\
    \includegraphics{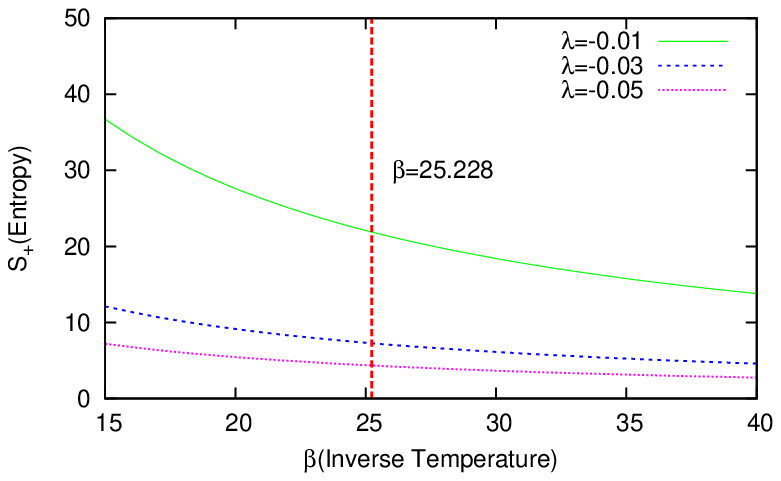}
  \caption{The figure illustrates the behavior of entropy of massless scalar wave from BRBH with $\lambda$ from $-0.1$ to $ 0$ at inverse Hawking temperature $\beta=25.228$, $\alpha=0.1$, $k_B=1$. The bottom picture shows the entropy together at different Hawking temperature for $\lambda=-0.01$ ,$ -0.03$, $-0.05$, respectively. }\label{fig2}
\end{figure}
\begin{figure}
  \centering
  % Requires \usepackage{graphicx}
  \includegraphics{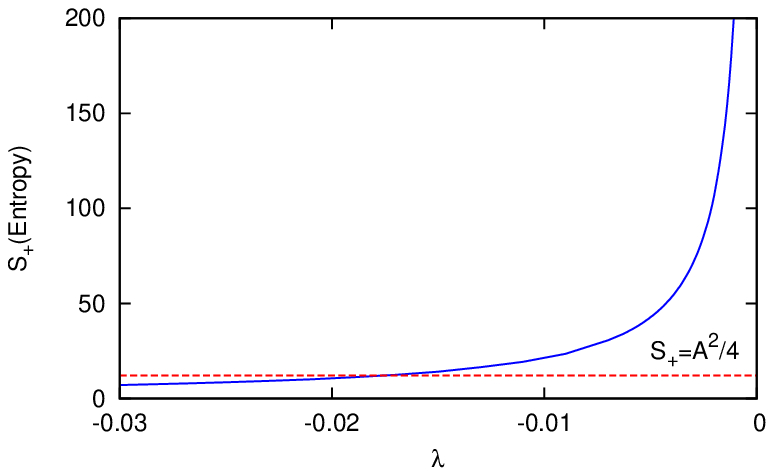}\\
    \includegraphics{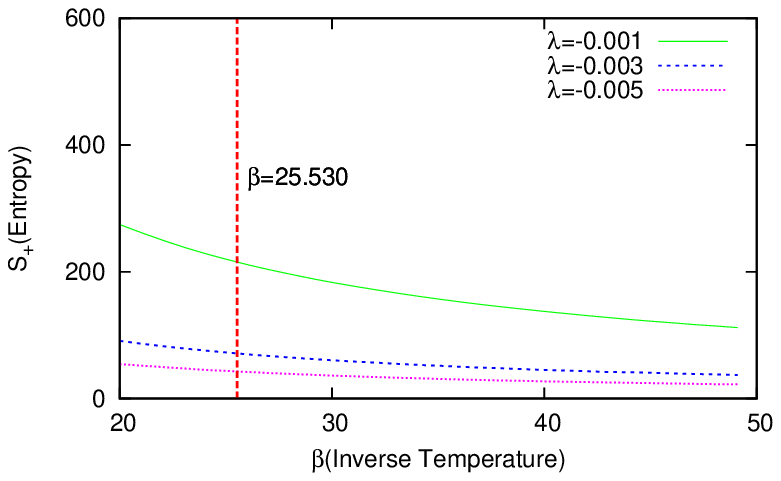}
  \caption{The figure illustrates the behavior of entropy of massless scalar wave from BRBH with $\lambda$ from $-0.03$ to $ 0$ at inverse Hawking temperature $\beta=25.530$, $\alpha=0.2$, $k_B=1$. The bottom picture shows the entropy together at different Hawking temperature for $\lambda=-0.01$ ,$ -0.03$, $-0.05$, respectively. }\label{fig3}
\end{figure}
\begin{figure}
  \centering
  % Requires \usepackage{graphicx}
   \includegraphics{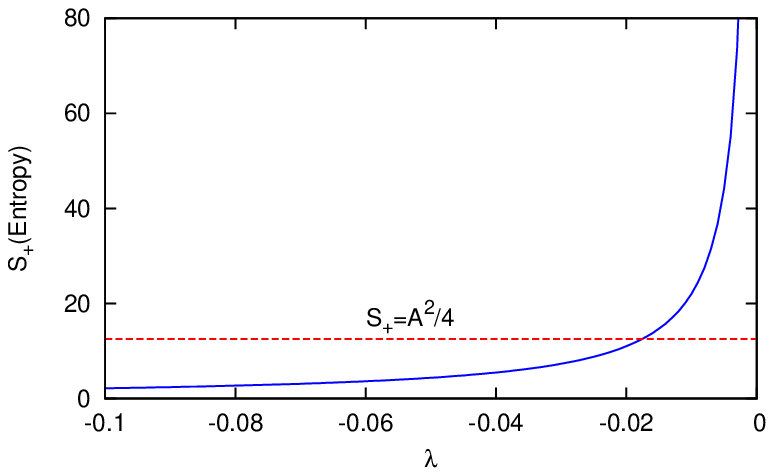}\\
    \includegraphics{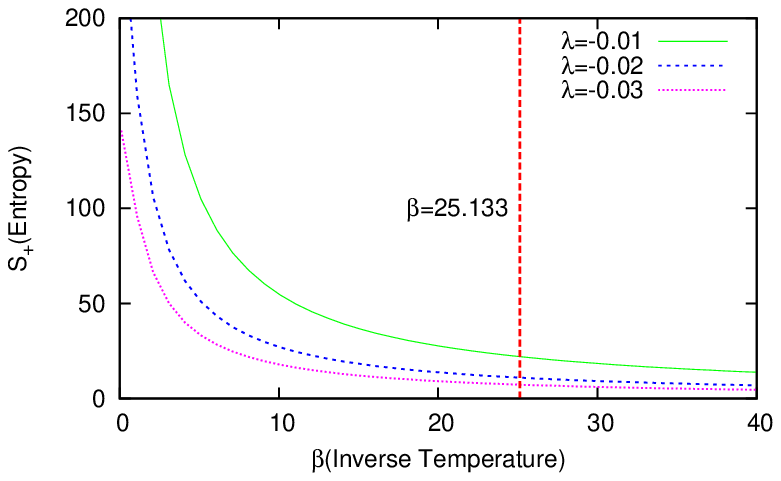}
  \caption{The figure illustrates the behavior of entropy of massless scalar wave from BRBH with $\lambda$ from $-0.1$ to $ 0$ at inverse Hawking temperature $\beta=25.133$, $\alpha=0$, $k_B=1$. The bottom picture shows the entropy together at different Hawking temperature for $\lambda=-0.01$ ,$ -0.03$, $-0.05$, respectively. }\label{fig4}
\end{figure}

Figs.\ref{fig2}-\ref{fig4} illustrate the behavior of the BRBH entropy with respect to the monopole charge $\alpha=0,0.1,0.2$ and parameter $\lambda$, inverse temperature. We assume $\lambda=\lambda l_p^2$, where $l_p$ is the Planck length, and in the system of the Planck units $l_p=1$. At the same temperature, the larger the value of $\lambda$, the larger the corresponding value of entropy is. When $\lambda$ tend to zero, the entropy increase rapidly. The value of $\lambda$ corresponding to actual entropy is approximately -0.02. When we used fixed value of $\lambda$, we find the parameter $\lambda$ makes the entropy lower.

In order to inspect the contribution of different parts of radius inside the external horizon, we maintain the entropy as $A/4$, and the radius is divided into 100 parts and integral. In Fig.5, we plot the integral of different parts for $\alpha=0$, we find the integral increases rapidly when the interval tends to external horizon. When the interval in the vicinity of zero, it makes little contribution to the entropy. We show the value of integral for different $\alpha$ in Table 1, we can see the last parts of integral constitute the main part of entropy, $\lambda$ approximately equivalent for different values of temperature. In Fig.6, we show the integral in logarithm forms to inspect more clearly. When $\alpha=0$(Schwarzschild case), there is no inner horizon, the entropy increases as integral parts tend to the external horizon, when $\alpha\neq0$, there are two extremum. It's obvious that the small vicinity of horizon radius show a great contribution than the parts around horizon radius and the contribution of the vicinity of external horizon almost takes the whole parts of the entropy. By comparing these curves, we can see that, the monopole charge $\alpha$ makes the entropy weaker and its distribution width narrower between two horizon. The inner horizon entropy  increases while the external horizon entropy decrease. Here we adopted the proper value of $\lambda$, and no cut-off is adopted.

In Fig.7, we try to discuss the situation outside horizon, it shows the contribution of vacuum outside the horizon. We plot the integral of entropy from external horizon $r=r_H$ to radius $r=100r_H$ for $\alpha=0$(Schwarzschild case),$\lambda=0.0175372$. When $r<10r_H$, the entropy maintains basically stable, when $r>20r_H$, the entropy increase as exponential form. This property indicates that if we make a cut-off between $r_H$ and $r_H+\varepsilon$($\varepsilon$ is a small quantity), we can get the Bekenstein-Hawking entropy. This agrees with the traditional brick-wall model\cite{12} and thin-layer model\cite{13}. In Fig.8, if we do not maintain $\lambda$ as a constant, the integral of entropy from $r_H$ to different upper limit can be adjusted to get the same Bekenstein-Hawking entropy.
\begin{figure}
  \centering
  % Requires \usepackage{graphicx}
    \includegraphics{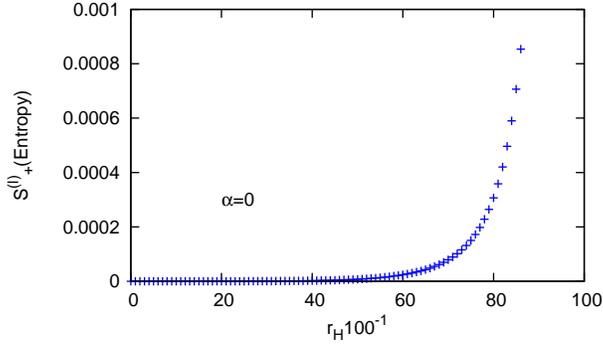}
  \caption{The integral of entropy of first 90 parts($0\rightarrow\frac{90}{100}r_H$) when $r_H$ is divided into 100 parts. The last 10 parts is showed in the table below.}\label{fig5}
\end{figure}

\begin{table}
\caption{ Entropy Integral between Different Parts of Radius.}

\begin{center}

 \begin{tabular}{|c|c|c|c|c|}
 \hline \hline
 &$\alpha=0$&$\alpha=0.1$&$\alpha=0.2$&$\alpha=0.3$\\
  \hline
 $\lambda$&-0.0175372&-0.0175376&-0.0174500&-0.0174500\\
$\beta$&25.1327412&25.2282438&25.5307184&26.0959399\\
$S_+$&12.5664&12.4719&12.1855&11.6975\\ \hline

 91&0.00205671&0.00204622&0.00201396&0.00195699\\
92&0.00268403&0.00266955&0.00262509&0.00254671\\
93&0.00360017&0.00357973&0.00351704&0.00340665\\
94&0.00500378&0.00497400&0.00488275&0.00472224\\
95&0.00729502&0.00724965&0.00711080&0.00686681\\
96&0.01137950&0.01130580&0.01108040&0.01068470\\
97&0.01967990&0.01954740&0.01914320&0.01843340\\
98&0.04065840&0.04037460&0.03951080&0.03799340\\
99&0.12414800&0.12325100&0.12053600&0.11575000\\
100&12.2699000&12.1775000&11.9590000&11.481000\\
 \hline
  \end{tabular}
   \end{center}
   \end{table}

\begin{figure}
  \centering
  % Requires \usepackage{graphicx}
    \includegraphics{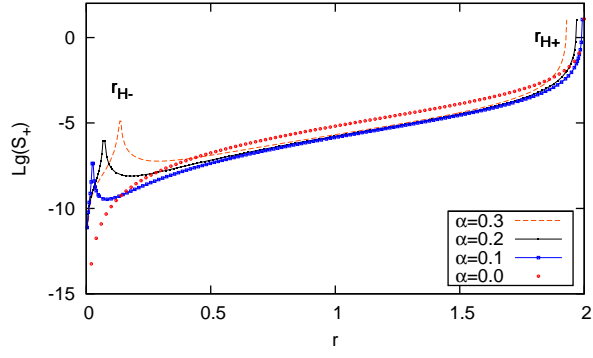}
  \caption{The logarithm of the integral of entropy for different interval within the external horizon $r_H$ for $\alpha=0$(Schwarzschild case),$ 0.1$, $0.2$, $0.3$ , respectively.}\label{fig6}
\end{figure}

\begin{figure}
  \centering
  % Requires \usepackage{graphicx}
    \includegraphics{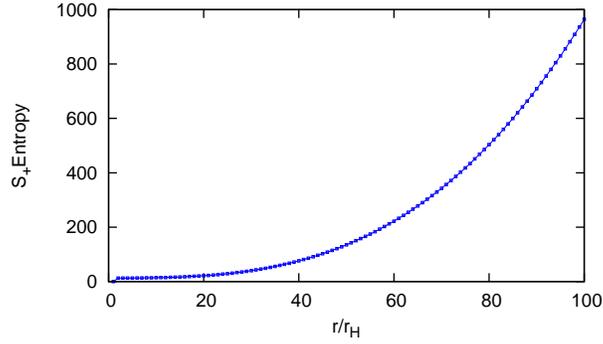}
  \caption{The integral of entropy from external horizon $r=r_H$ to  radius $r=100r_H$ for $\alpha=0$(Schwarzschild case),$\lambda=0.0175372$. It shows the contribution of vacuum outside the horizon. }\label{fig7}
\end{figure}
\begin{figure}
  \centering
  % Requires \usepackage{graphicx}
    \includegraphics{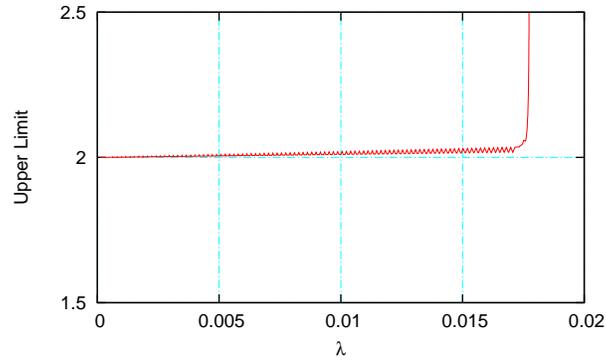}
  \caption{The integral upper limit corresponding to different value of $\lambda$ when entropy is maintain as $A/4$.}\label{fig8}
\end{figure}

In conclusion, we have investigated the entropy of scalar field on a BRBH backgrounds using analytic and numerical methods. We found the monopole charge $\alpha$ makes the entropy weaker, and the distribution width of entropy between two horizons become narrower. We gave the distribution of the entropy in the interior of BH, we also found that the contribution of the vicinity of external horizon almost taken the whole parts of entropy. This calculation did not need to introduce any cut-off while considered the exterior region of horizon we need to cut out a small region to get the  Bekenstein-Hawking entropy, this agreed with the previous works list in the introduction.

This project is supported by the National Natural Science Foundation
of China under Grant No.10873004, the State Key Development Program
for Basic Research Program of China under Grant No.2010CB832803 and
the Program for Changjiang Scholars and Innovative Research Team in
University, No. IRT0964.

%\newpage


\begin{thebibliography}{99}
\bibitem{1} J. D. Bekenstein, Lett. Nuovo Cimento 4 737(1972);
J. D. Bekenstein, Phys. Rev. D 7 2333(1973);
J. D. Bekenstein,  Phys. Rev.D 9 3292(1974);
 S. W. Hawking, Commun. Math. Phys. 43 199(1975).
\bibitem{2}G. 't Hooft, Nucl. Phys. B256, 727 (1985).
\bibitem{3} X. Li and Z. Zhao, Mod. Phys. Letts. A 15 1739(2000);
 X. Li and Z. Zhao,  Int. J. Theor. Phys. 40 90(2001);
T. P. Song, C. X. Hou and W. L. Shi, Acta Phys. Sin. 51 06(2002).
\bibitem{4} D. J. Gross and P. F. Menda   Nucl. Phys. B 303 407(1988).
\bibitem{5} Y. W. Kim and Y. J. Park, Phys. Rev. D 77 067501(2008).
\bibitem{6}  M. Yoon, J. Ha and W. Kim, Phys. Rev. D 76 047501(2007).
\bibitem{7} K. Nouicer, Phys. Lett. B 646, 63(2007).
\bibitem{7.1}G. Amelino-Camelia, M. Arzano, Y. Ling, and G.Mandanici, Classical Quantum Gravity 23, 2585 (2006)
\bibitem{7.2}S. Hossenfelder, Phys. Rev. D 73, 105013 (2006);
S. Hossenfelder,Class. Quantum Grav. 23, 1815 (2006).
\bibitem{7.3}M. Fontanini, E. Spallucci, and T. Padmanabhan, Phys.Lett. B 633, 627 (2006).
\bibitem{8} L. N. Chang, D. Minic, N. Okamura, and T. Takeuchi,Phys. Rev. D 65, 125028(2002).
\bibitem{9} Y.W. Kim, Phys. Rev. D 77, 067501(2008).
\bibitem{10} X. Li, Phys. Lett. B 540, 9(2002).
\bibitem{11} X. Sun and W. Liu, Mod. Phys. Lett. A 19 677(2004).
\bibitem{12}  R. Zhao,  Y. Q. Wu and  L. C. Zhang Class. Quantum Grav. 20 4885(2003)
 C. Liu,  X. Li and Z. Zhao, Gen. Rel. Grav. 36 1135(2004);
C. Z. Liu, Int. J. Theor. Phys. 44 567(2005).
\bibitem{13}  W. Kim, Y. W. Kim and Y. J. Park Phys. Rev. D 74 104001(2006);
 W. Kim,  Y. W. Kim and  Y. J. Park, Phys. Rev. D 75 127501(2007)
\bibitem{14} W. B. Liu, Chin. Phys. Lett. 20 440(2003)

\end{thebibliography}
\end{document}